\documentclass{aa}

\usepackage{graphicx,epsfig,fancyhdr,psfig,rotating,amsmath,natbib}
\def\kms{\rm km\;s$^{-1}$}

\begin{document}

\title{Kinematics and helicity evolution of a loop-like eruptive prominence }

\author{K. Koleva
	\inst{1}
\and  M.S. Madjarska\inst{2}
\and
       P. Duchlev \inst {1}
\and
        C. J.~Schrijver\inst{5}
\and
	J.-C. Vial\inst{3,4}
\and
	E. Buchlin\inst{3,4}
\and
M. Dechev \inst {1}
	}

\offprints{koleva@astro.bas.bg}

\institute{Institute of Astronomy and National Astronomical Observatory,\\
 Bulgarian Academy of Sciences, 72 Tsarigradsko Chaussee Blvd., 1784 Sofia, Bulgaria
 \and
	Armagh Observatory, College Hill, Armagh BT61 9DG, N. Ireland
\and
	CNRS, Institut d'Astrophysique Spatiale, UMR8617, 91405 Orsay, France
\and
        Univ Paris-Sud, Institut d'Astrophysique Spatiale, UMR8617, 91405 Orsay, France
\and
 Solar and Astrophysics Lab., Lockheed Martin Advanced Techn. Ctr.,
3251 Hanover St., Bldg. 252, Palo Alto, CA 94304-1191, USA}

 \date{Received date, accepted date}

\abstract
{}
{We aim at investigating the morphology, kinematic and helicity evolution of a loop-like prominence during its eruption.}
{We use multi-instrument observations from AIA/SDO,  EUVI/STEREO and LASCO/SoHO.
The kinematic, morphological, geometrical, and helicity evolution of a  loop-like eruptive prominence are studied in the context of the magnetic flux rope model of solar prominences. }
{The prominence eruption evolved as a height expanding twisted loop with both legs anchored in the chromosphere of a plage area.
The eruption process consists of  a prominence activation,  acceleration, and a phase of constant velocity. The prominence body was composed of  left-hand (counter-clockwise) twisted threads around the main prominence axis. The twist  during the eruption was estimated at $6\pi$ (3 turns). The prominence reached a maximum height of 526 Mm before contracting to its primary location and partially reformed in the same place  two days after the eruption.  This ejection, however, triggered a CME seen in LASCO C2. The prominence was located in the northern periphery of the CME magnetic field configuration and, therefore,  the background magnetic field was asymmetric with respect to  the filament position. The physical conditions of the falling plasma blobs were analysed with respect to the prominence kinematics.}
{The same sign of the prominence body twist and writhe, as well as the amount of twisting above the critical value of $2\pi$ after the activation phase indicate that possibly  conditions for kink instability were present. No signature of magnetic reconnection was observed anywhere in the prominence body and its surroundings. The  filament/prominence descent following the eruption and its  partial reformation at the same place two days later suggest a  confined type of eruption.  The asymmetric background magnetic field  possibly played  an important role in the failed eruption.}

\authorrunning{K. Koleva et al.}
\titlerunning{Kinematics and helicity evolution}

\maketitle

\section{Introduction}

Prominence eruptions are large-scale eruptive phenomena which occur  in the low solar atmosphere. Observations show that prominences display a wide range of eruptive activity. There are three types of prominence (filament) eruptions according to the observational definitions of \citet{2007SoPh..245..287G} based on the relation between the filament mass and corresponding supporting magnetic structure: full, partial, and failed (confined), of which the partial ones are the most complex.
A full eruption occurs when the entire magnetic structure and the pre-eruptive prominence material are expelled into the heliosphere. The case when neither the filament mass, nor the supporting magnetic structure escape the solar gravitational field is a failed eruption. Partial eruptions can be divided into two subcategories: i) when the entire magnetic structure erupts, with the eruption containing either part or none of its supported pre-eruptive prominence material, and ii) when the magnetic structure itself partially escapes with either some or none of the filament mass \citep{2007SoPh..245..287G}. One important observational consequence concerning partial and failed eruptions is the re-formation of the filament at the pre-eruptive location.

\citet{2004ApJ...602.1024S, 2004ApJ...613.1221S} unveiled  a common pattern of prominence eruptions: an initial slow-rise phase (with a very small acceleration), during which the filament gradually ascends, followed by a sharp change to a phase of fast acceleration. There exist three types of prominence eruption after the fast rise phase: i) an eruptive prominence can continue to rise with acceleration, ii) the fast rise can be followed by a constant velocity phase, or iii) the constant velocity phase of an eruptive prominence can be followed by a deceleration phase \citep{1998ASPC..150..302V}.

Eruptive prominences (EPs) (or filaments if observed on the solar disk) are frequently associated and physically related to coronal mass ejections (CMEs) and flares \citep{1995ASSL..199.....T, 1976SoPh...48..159W, 1979SoPh...61..201M, 1987SoPh..108..383W, 1991SoPh..136..379S}. Usually, all three eruptive events occur in the same large-scale coronal magnetic field, in which the EP only occupies a limited volume at its base. Such an association  suggests  that possibly the same physical process drives these eruptive phenomena. Filament-CME relationship studies often do not take into account the relation between filament-related magnetic field configurations, like for instance,  an empty filament channel, and a CME initiation \citep{2006SSRv..123...81A}. Therefore, the  filament-CME relationship can be even stronger than is known so far. An EP and/or a CME are considered as the eruption of a preexisting magnetic flux rope (MFR) or an initial magnetic arcade that evolves into a magnetic flux rope during the eruption process via magnetic reconnection \citep{2006ApJ...649..452C, 1999ApJ...510..485A}. However, the question whether an MFR exists prior to the eruption  or it is formed during the eruption  remains controversial, with the MFR topology preceding the eruption  being favoured in many studies during the last decade.

During activation, the slow rise motion of an EP is usually accompanied by a gradual morphological evolution from an initially intricate structure into an apparently toroidal shape, exposing sometimes a twisted pattern, that is most often prominent in the legs of the prominence \citep{1991SoPh..136..151V}. EPs very often develop a clearly helical shape in the course of the eruption, which is the characteristic signature of a  MagnetoHydroDynamic   (MHD) kink instability  of a twisted MFR \citep[e.g.][]{2005ApJ...622L..69R}. A MFR becomes kink-unstable if the twist exceeds a critical value of  2$\pi$ \citep[e.g.][]{1981GApFD..17..297H, 2005ApJ...630..543F, 2005ApJ...630L..97T}. The axis of an MFR then undergoes writhing (kinking) motions and part of the twist of the field is transformed into helical writhe of the axis, since the magnetic helicity is conserved \citep{2005ApJ...622L..69R}. The conservation of helicity in ideal MHD \citep{1984GApFD..30...79B} requires the resulting writhe to be of the same sign as the transformed twist. The kink instability has long been investigated as a possible triggering mechanism for solar eruptive phenomena, especially in flux rope models \citep{2009ApJ...697..999L}.

Kink and torus instabilities are suggested to be two mechanisms for triggering solar flares and coronal mass ejections \citep{1976PASJ...28..177S, 2005ApJ...630L..97T, 2006PhRvL..96y5002K, 2008ApJ...674..586S}.
Kink instability was first suggested as the trigger of prominence eruptions (confined and ejective) by \citet{1976PASJ...28..177S} but has been generally regarded as a possible explanation only for confined events \citep[e.g.][]{2003SoPh..214..151G}. \citet{2005ApJ...630L..97T}, and \citet{2005ApJ...630..543F} succeeded in modelling full ejection of a coronal flux rope from the Sun driven by kink instability.
A  kinking may be an important factor in the erupting process, but the type of eruption may strongly depend on the role played by magnetic  reconnection and its location with regards to the prominence body \citep{1999ApJ...510..485A}. As discussed in \citet[][and the references therein] {2007SoPh..245..287G} the kink instability as a main trigger of prominence destabilization and eruption is challenging  to be proven observationally because the helicity can force a flux rope to  writhe, without any instability occurring.

\citet{2006PhRvL..96y5002K} studied the expansion instability of a toroidal current ring embedded  in a low-beta magnetised plasma. \citet{2008ApJ...674..586S} analysed  two near-limb filament destabilisations involved in CMEs. Numerical simulation of a torus instability showed a relatively close quantitative match of the observations, implying that these two cases are likely to be torus instability eruptions.

Observations of filament eruptions that strongly suggest a helical kink occurring in flux rope topology were presented by \citet{2005ApJ...628L.163W} for  a full eruption, by \citet{2006ApJ...651.1238Z} and by \citet{2007ApJ...661.1260L} for partial eruptions, and \citet{2003ApJ...595L.135J} and \citet{2006ApJ...653..719A} for confined (failed) eruption. An observational definition of kinking related to the different types of filament eruptions is given by \citet{2007SoPh..245..287G} and \citet{2007SoPh..246..365G}.

In this paper, we investigate  the phenomenology of an eruptive  loop-shaped helically twisted prominence  with fixed footpoints using state-of-art observations from the Atmospheric Imaging Assembly (AIA) aboard the Solar Dynamics Observatory (SDO) in the 304~\AA\ EUV passband. We have the unique opportunity to combine limb with on-disk observations of an eruptive prominence thanks to the EUVI/STEREO B observations which at the time of the observations was at an angular distance of 71 degrees with the Earth. We aim at investigating the morphology, kinematic and helicity evolution of a loop-like prominence during its eruption. In Section 2 we describe the set of observations used in this study. In Section 3 we present the main results, which are discussed in Section 4. The conclusions are drawn in Section 5.

\begin{figure}[!ht]
\begin{center}
\includegraphics [scale=0.9]{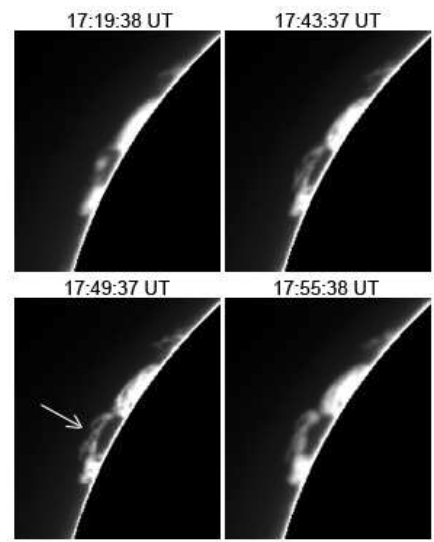}
\caption{Mauna Loa H$\alpha$ images before and during the start of the prominence activation. The arrow points at the region revealing twisted structure.}
\label{fig1}
\end{center}
\end{figure}

\section{Observations and data analysis}
\label{sect2}

The prominence eruption occurred at the North-East solar limb between 17:30~UT and 19:30~UT on 2010 March 30. The EP was centered at mean heliographic co-ordinates N$22.63^{\circ}$; E$78.80^{\circ}$  and mean position angle $66^{\circ}$.

\begin{figure*}[!ht]
\begin{center}
\vspace{-4cm}
\hspace{-4cm}
\includegraphics[width=\textwidth]{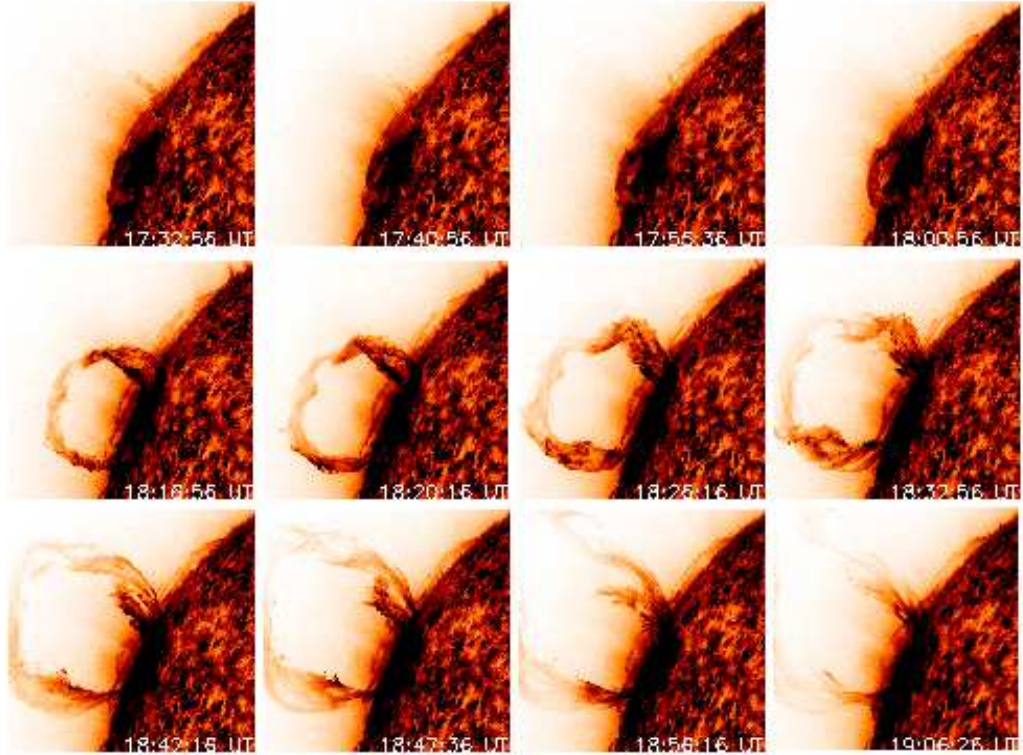}
\caption {He~{\sc ii}~304~\AA\ AIA images (in reversed colour table)  showing the morphology and, in particular, the helicity evolution of the erupting prominence. AIA He~{\sc ii} 304~\AA\ image animation of the prominence eruption can be seen online.}
\label{fig2}
\end{center}
\end{figure*}

For the present study we used  images taken with 1~min cadence in the He~{\sc ii}~304~\AA\ passband of AIA/SDO \citep[AIA;][]{2011SoPh..tmp..172L}. The AIA consists of seven Extreme Ultra-Violet (EUV) and three Ultra-Violet (UV) channels which provide an unprecedented view of the solar corona with an average  cadence of $\sim$12~s. The AIA image field-of-view reaches  1.3 solar radii with a  spatial resolution of $\sim$1.5\arcsec. We used level 1 reduced data, i.e. with the dark current removed and the flat-field correction applied.  The images were further stabilised  for the satellite movements by applying intensity cross-correlation analysis  using the SolarSoftware procedure get\_correl\_offsets.pro.  For the purpose of our analysis we defined the visible limb from the  AIA He~{\sc ii}~304~\AA\ images.  Note that the jitter correction was only needed this early  in the mission as the image stabilisation was still subject to calibration in the commissioning phase when these observations were taken.

We also analysed observations from the Extreme Ultraviolet Imager (EUVI) aboard STEREO Behind (B) spacecraft. EUVI has a field-of-view  of 1.7$R_{\odot}$ and observes in four spectral channels (He~{\sc ii}~304~\AA, Fe~{\sc ix/x}~171~\AA, Fe~{\sc xii}~195~\AA\ and Fe~{XIV}~284~\AA) that cover the 0.1 to 20~MK temperature range \citep{2004SPIE.5171..111W}. The EUVI detector has  2048~$\times$~2048 pixels$^2$ size and a pixel size of 1.6\arcsec.
In the present study we used images in the He~{\sc ii}~304~\AA\ channel with an average cadence of 10 minutes.
Images obtained by the Large Angle and Spectrometric Coronagraph (LASCO)/C2 on board SOHO, whose field-of-view extends from 2 to 6 solar radii \citep{1995SoPh..162..357B} were also analysed in the present study.

\begin{figure*}[!ht]
\begin{center}
\vspace{-2cm}
\includegraphics[scale=0.9]{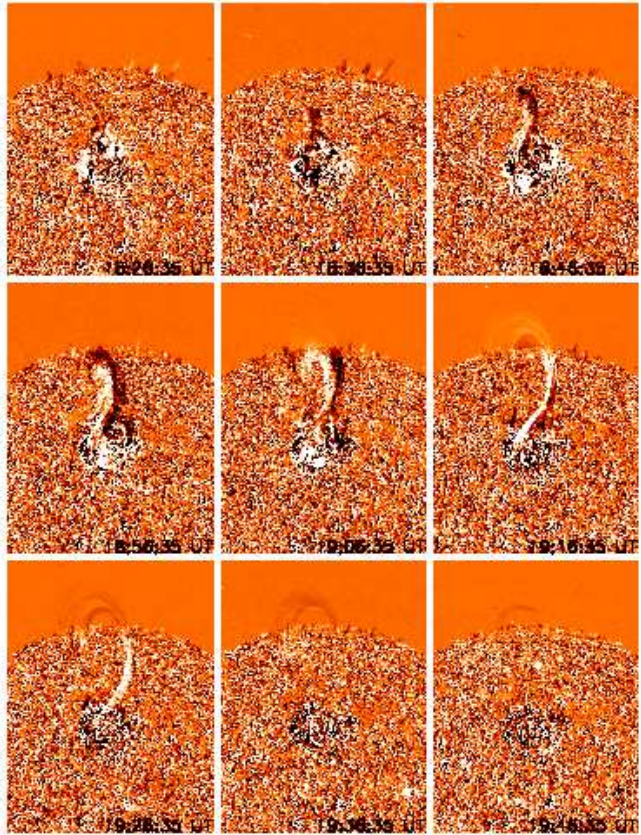}
\vspace{0cm}
\caption{Running difference images from EUVI B in the  He~{\sc ii}~304~\AA\ channel. EUVI/STEREO B He~{\sc ii} 304~\AA\ image difference animation of the prominence eruption on 2010 March 30 can be seen online.}
\label{fig3}
\end{center}
\end{figure*}

\begin{figure*}[!ht]
\begin{center}
\includegraphics[scale=0.7]{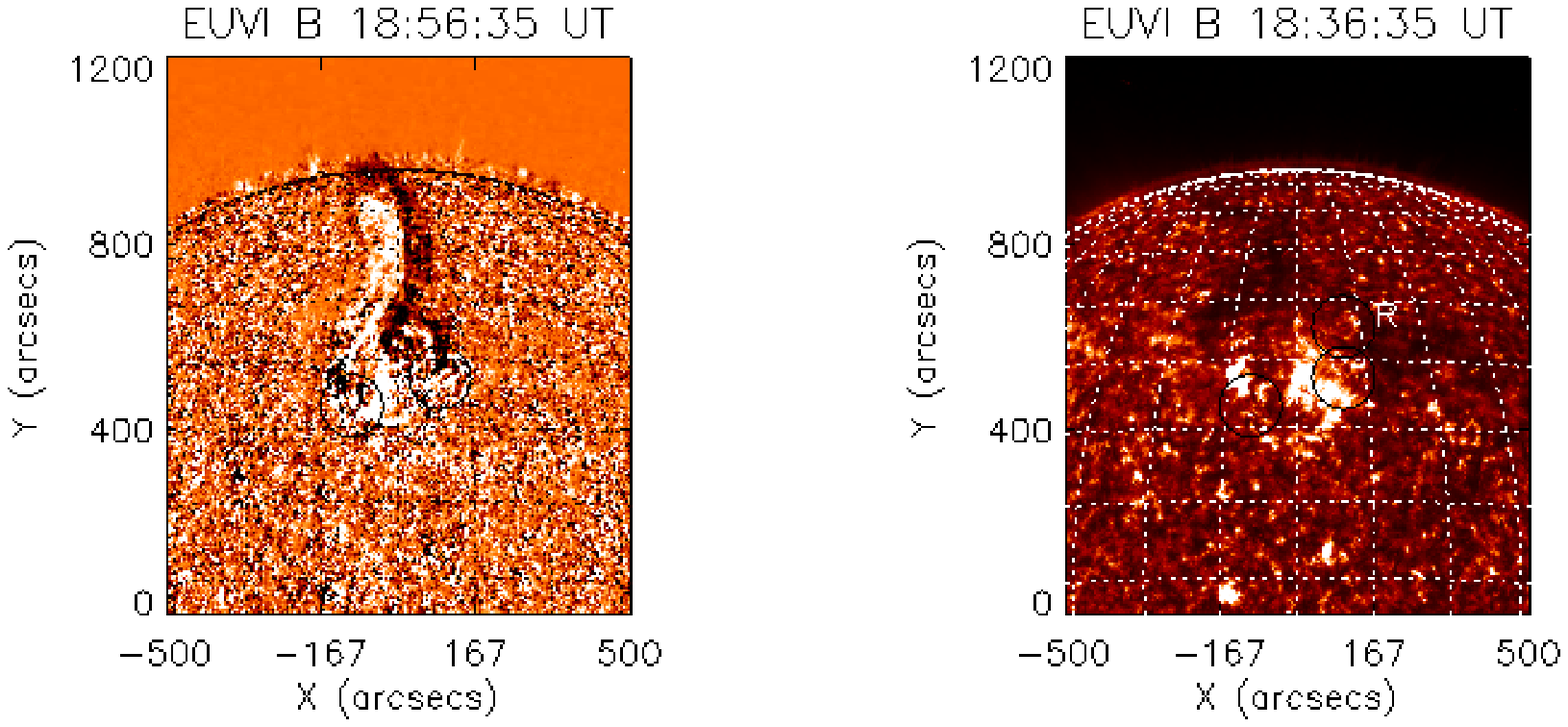}
\includegraphics[scale=0.7]{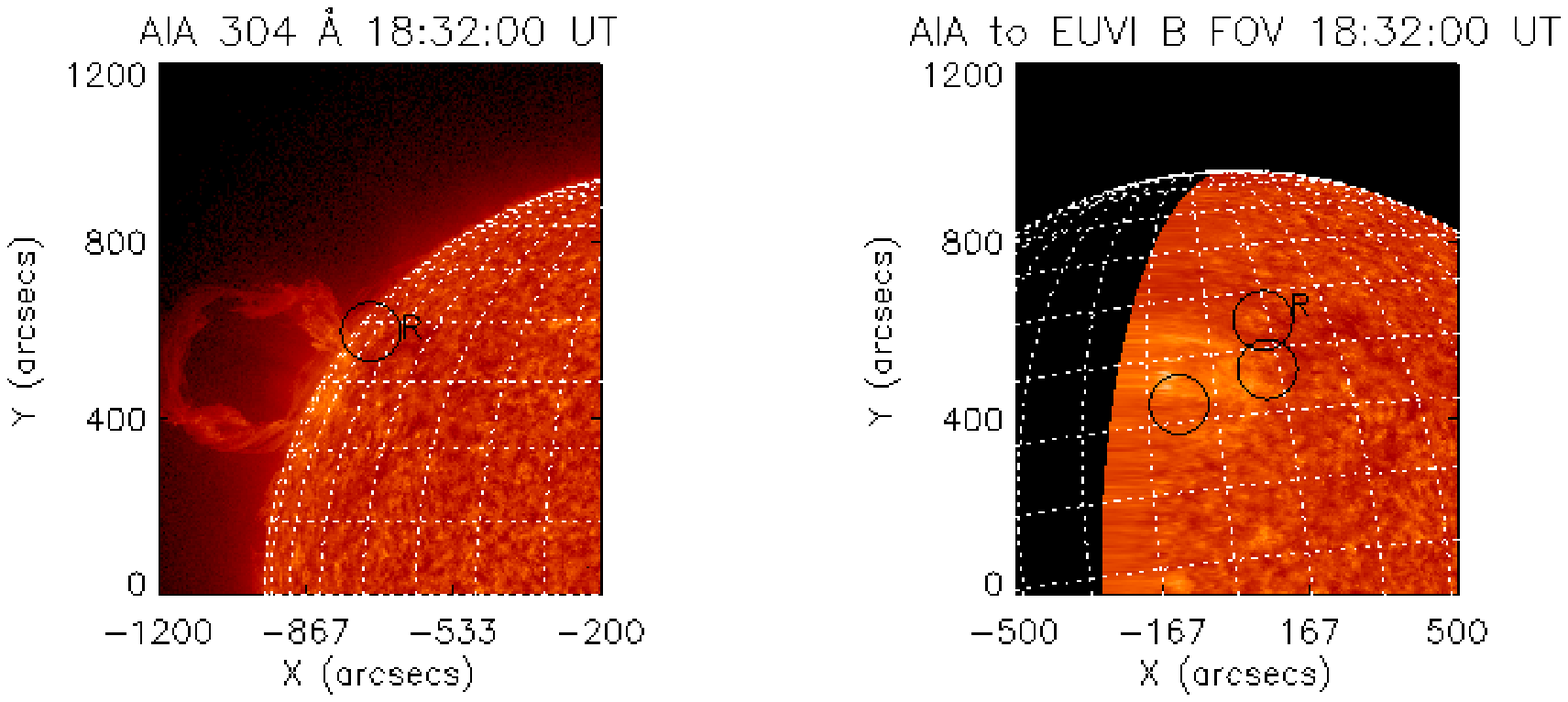}
\caption{EUVI difference image (top left) and EUVI image (top right). AIA image (bottom left) and the same image transformed to the STEREO FOV (bottom right). The filament's footpoints and  a reference feature `R' above the northern leg are marked with circles.}
\label{fig4}
\end{center}
\end{figure*}

\section{Results}
\label{new}

\subsection{Morphology and helicity evolution}

\begin{table*}
\centering
\caption{The kinematics of the eruptive prominence derived from the AIA images. }
\label{T1}
\begin{tabular}{l c  c c c}
\hline\hline
Phase & Time (UT) &  Height (Mm) & Velocity (\kms) & Acceleration (m~s$^{-2}$)\\
\hline\hline
Activation& 17:33 -- 18:00 &  18 - 34 & 10 & 0 \\
Acceleration& 18:00 -- 18:12 &  34 - 121 & 15 -- 166 &46 -- 430 \\
Constant velocity &18:12 -- 18:44 & 121 -- 295 & 91 & 0 \\
\hline
\end{tabular}
\end{table*}

The  prominence eruption evolved  as a height expanding  twisted loop  with both legs anchored in the chromosphere of a plage area.
The prominence  before the activation can already be seen in H$\alpha$  at 17:19~UT  (Fig.~\ref{fig1}). It  shows as a few bright clouds above the limb with no distinguishable fine structure. In the AIA He~{\sc ii} 304~\AA\ images at 17:32~UT and 17:40~UT  in Fig.~\ref{fig2} (see  also the online material), the fine structure of the prominence appears very dense, the prominence lies very close to the limb  and, therefore,  it is impossible to distinguish its small-scale structure.  Note that in H$\alpha$ we see the cold core of the prominence, while in He~{\sc ii}~304~\AA\ it is the envelope at higher temperatures \citep{2010SSRv..151..243L}.
The H$\alpha$ image  at 17:49~UT shows the first signature of a twisted prominence fine structure.
From part of the prominence where the fine structure is visible, we could estimate a $2\pi$ twist, i.e.  one turn of the prominence rope around  its main axis. This time coincides with the beginning  of the prominence slow rise during the activation phase. The full amount  of  prominence twist is first seen at around 18:20~UT  (Fig.~\ref{fig2}) in the AIA He~{\sc ii}~304~\AA\ images. We estimated from  the AIA 304~\AA\  images at 18:20~UT, 18:26~UT and 18:32~UT,  a total twist of about $6\pi$ (3 turns) of the  eruptive prominence body.

 We also searched for a signature of magnetic reconnection in all available data including AIA and EUVI He~{\sc ii} 304~\AA\ and Fe~{\sc xii} 195~\AA. Both the off-limb AIA and on-disk EUVI data do not show any brightening in  the footpoints of the prominence, their surroundings as well as along the prominence body. This suggests that magnetic reconnection may not have had contributed to the prominence destabilization and  eruption.

\begin{figure*}[!ht]
\begin{center}
\includegraphics [scale=0.8]{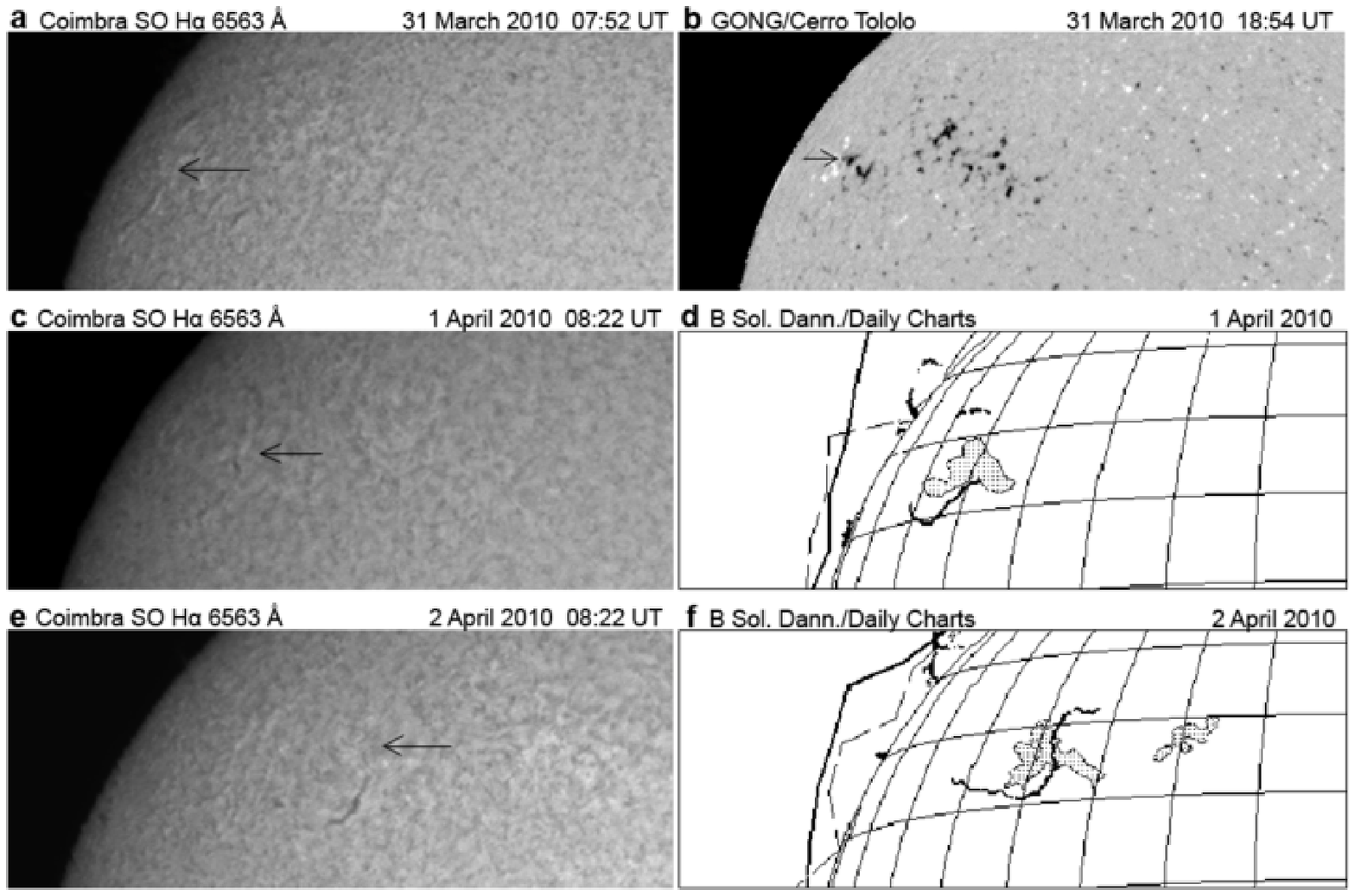}
\caption{ The prominence reformation  as seen on the disk as a filament  traced in Coimbra Solar Observatory  $H_{\alpha}$ images (\textbf{a, c, e}), a  GONG magnetogram (\textbf{b}), and  daily drawings of the Sun from the electronic Bulletin "Solnechnye Dannye" of Pulkovo Observatory (\textbf{d} and \textbf{f}).}
\label{fig5}
\end{center}
\end{figure*}
After 18:25~UT the eruption can also be followed  on the EUVI/STEREO image differences  shown in Fig.~\ref{fig3} (see also the online material). Here, thanks to  the different view point (see the AIA view translated on the EUVI B image in Fig.~\ref{fig4})  we could see that the filament upper body underwent a right-hand writhe.  The deformation of the prominence  is  still not visible in the AIA images at this time (Fig.~\ref{fig2}) due to the line-of-sight effect.  At 18:56~UT   the crossing  of the prominence body axis due to the writhe  as seen in projection on the disk, in the STEREO difference images (Fig.~\ref{fig3}) is visible at approximately half of the rising filament height. At 19:06~UT, this crossing  is visible at approximately one third from the feet of the prominence. The writhe increased with time, reaching $\pi$/2 at 19:26~UT (Fig.~\ref{fig3}).

The  AIA He~{\sc ii} 304~\AA\ images after 18:26~UT reveal best the helical twist of the prominence and its evolution in time (last row of Fig.~\ref{fig2}).  We can clearly see that  during the prominence uplift the twist  transfers from the lower (legs)  to  the upper prominence body  causing the prominence to evolve from a loop-shaped into a ribbon-like structure. With the help of  the EUVI B difference images (Fig.~\ref{fig3}), we  established the footpoint location  noted with black circles in Fig.~\ref{fig4} (top left). The so defined foot positions are also shown on the EUVI image  at 18:36~UT (Fig.~\ref{fig4}, top right). A feature above the northern leg is also encircled and labelled with `R'. This feature is used as a reference for the translation of the AIA images into the STEREO B view. In Fig.~\ref{fig4} (bottom right) we show the AIA~304~\AA\ image at 18:32~UT with the expanding loop-like prominence.  This image was transformed to the STEREO B view and the legs of  prominence identified in the EUVI B images were transposed (shown with black circles) as well as the reference feature.  We paid special attention to the localisation of the feet of the EP in order to determine the type of twist and writhe, and their evolution in time. Based on the magnetic field polarity configuration obtained from the magnetograms, the H$\alpha$ images and the prominence drawings (Fig.~\ref{fig5}), we could determine in which magnetic polarity the southern and the northern legs of the prominence were located.
From all the above information we established  that the prominence underwent a  writhe as a result of counter-clockwise rotation with respect to the two polarity fluxes.

\subsection{Kinematics}

\begin{figure}[!ht]
\begin{center}
\includegraphics [scale=0.9]{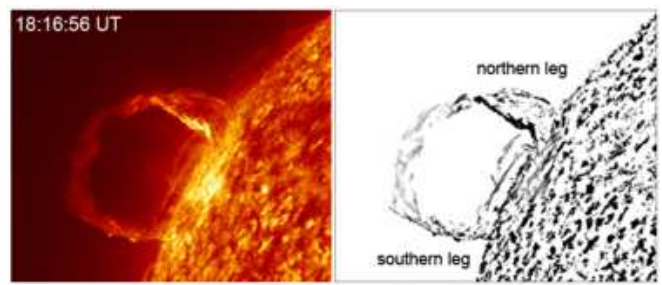}
\caption{The eruptive prominence on 2010 March 30 observed in the 304 \AA\ AIA/SDO channel at 18:17 UT ({\bf left}) and  its corresponding edge-enhanced image (\textbf{right}).}
\label{fig6}
\end{center}
\end{figure}

\begin{figure}[!ht]
\begin{center}
\includegraphics [scale=0.6]{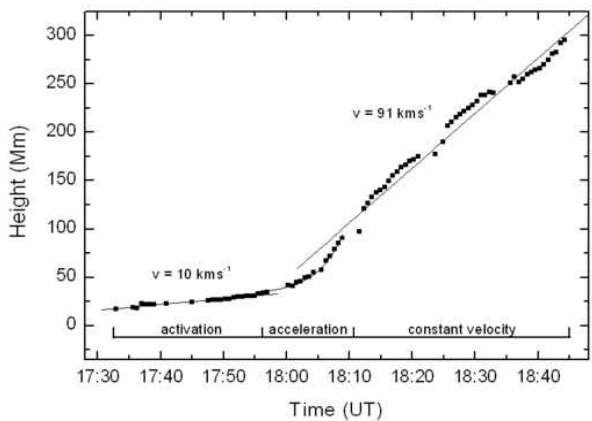}
\caption{Height-time profile of the prominence eruption for the three phases of the EP evolution:  the activation,  the acceleration, and the constant velocity.}
\label{fig7}
\end{center}
\end{figure}

The prominence height was determined as the height of the main axis of the prominence above the visible limb as observed in  the He~{\sc ii}~304~\AA\ channel of the AIA/SDO images (Fig.~\ref{fig6}).  The time evolution of the height reveals three distinctive phases of the prominence eruption: a  prominence activation, an eruption with  acceleration, and an eruption with a constant velocity (Fig.~\ref{fig7}). From the first and second derivatives of the polynomial fit of the height-time curve, we defined the speed and the acceleration of the prominence eruption. The prominence activation is already in progress at the begining of the AIA observations around 17:33~UT. It is defined as the time period  of a slow rise of the loop system with a velocity of 10~\kms. Until 18:00~UT the height of the prominence changed from 18~Mm to 34~Mm. The eruption onset was registered at 18:00 UT and it was determined from the AIA images as a sudden increase of the prominence height. It lasted until around 18:12~UT (Fig.~\ref{fig7}). During this phase  the prominence height changed from 34~Mm to 121~Mm and the speed of the prominence rise increased from 15~\kms\ to a maximum of 166~\kms\ with an acceleration from 46 to 430~m~s$^{-2}$. The constant velocity phase  was measured until 18:44~UT. After this time the top of the looped prominence is outside the AIA field-of-view.
The kinematics of the different phases of the eruptive process is summarised in Table~\ref{T1}.

 It took 80~min  the prominence to reach its maximum height (Fig.~\ref{fig3})  at 19:16~UT as seen in the EUVI B images (note that the top of the prominence quits the AIA FOV after 18:44~UT). We were able to measure a descent of the prominence body with a velocity of $\sim$50~\kms\  using three EUVI images  in which the prominence is clearly visible above the limb, i.e. at 19:16~UT, 19:26~UT and 19:36~UT (last row of Fig.~\ref{fig3}).

After 18:47~UT (Fig.~\ref{fig2}) the prominence legs are no longer twisted and we can clearly observe the downflow of individual plasma blobs thanks to the high AIA cadence of 1~min as well as its excellent spatial resolution. From the measurements of four blobs we found  an average downflow speed of 31.8~\kms\, with a minimum of 13.7~\kms\ and maximum of 60~\kms\ . Because of the continuous draining of the plasma towards the chromosphere,  the prominence plasma density becomes lower and the prominence  fainter  in the AIA and EUVI B He~{\sc ii} 304~\AA\ images.

\subsection{A CME association}

\begin{figure*}[!ht]
\begin{center}
\includegraphics [scale=1.0]{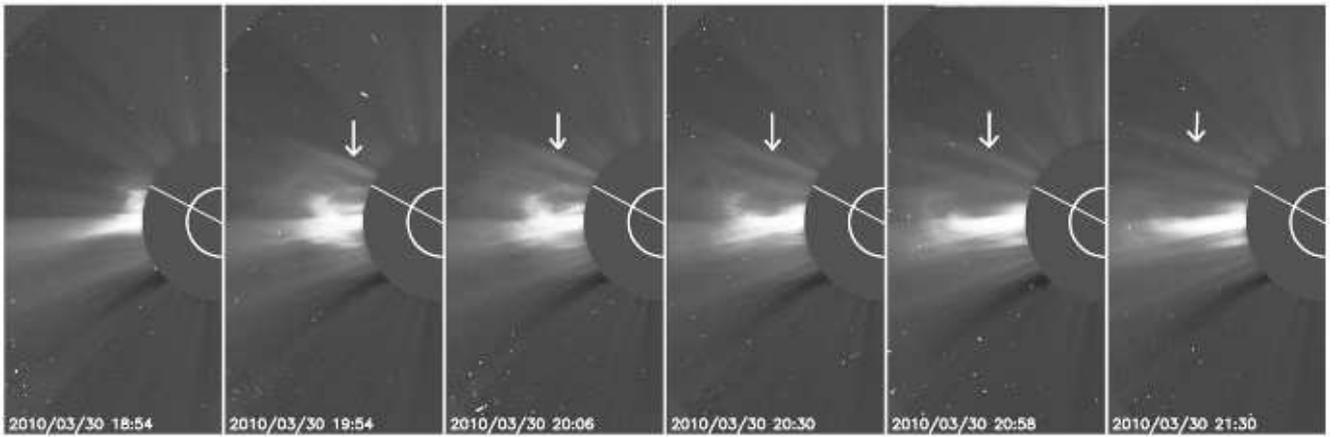}
\caption{SOHO/LASCO C2 registrations between 18:54 UT and 23:42 UT on 30 March 2010. The CME is first shown at 18:54 UT with a central position angle of $86^{\circ}$. The EP is located in the northern periphery of CME magnetic system. The white radial line over-plotted on the occulting disk indicates the central EP position angle ($55^{\circ}$) at the eastern limb.}
\label{fig8}
\end{center}
\end{figure*}

At 18:30 UT C2/LASCO  registered a CME at the eastern limb which was associated with the eruptive prominence.
The width of the CME was $64^{\circ}$ and the position angle was $74^{\circ}$. The prominence is located at the northern periphery of the CME magnetic system, at $19^{\circ}$ to the north of the position angle of the highest part of the CME. The CME propagated in the C2/LASCO field-of-view  with a velocity decreasing  from 853~\kms\ to 599~\kms, implying a deceleration of 16.4 m~s$^{-2}$. The  average linear velocity was 724~\kms.   At the location of the prominence  eruption an expanding loop  with an empty cavity underneath  which is part of the CME system is well seen in the C2/LASCO images (indicated with an arrow in Fig.~\ref{fig8}). Assuming  a constant velocity of 91~\kms\  for the prominence eruption, at 18:54~UT, the prominence had reached  a height of 390 Mm while at 19:16~UT, when it reached its maximum height, it was at 526~Mm (less than one solar radius).This strongly supports the observation that no prominence material is seen in the C2/LASCO FOV.

\subsection{Prominence reformation}

The prominence partially reformed in the same place  two days after its eruption.  In Fig.~\ref{fig5} the re-formation of the filament is presented from March 31 to  April 2. On  March 31, one day after its eruption, the filament can be traced only above and below the plage  area on the Coimbra Solar Observatory $H_{\alpha}$ (Fig.~\ref{fig5}a). The magnetic field of the plage area associated with the reforming filament can be traced  in Fig.~\ref{fig5}b.  Fig.~\ref{fig5} c, d, e and f
show H$\alpha$ images and drawings of the area from the Pulkovo Observatory, confirming the reformation of the filament. However, we can notice that the H$\alpha$ filament is rather
faint. We searched for its reformation in the EUV lines of EUVI B, but even in He~{\sc ii}
it was not possible  to detect it.  One explanation can be that some of the prominence material  may have set back  to the photosphere because of the draining (Sect.~3.2) leaving the remaining material to settle in the magnetic skeleton of the prominence in the corona. Another possibility is that  the material was not totally recovered at the low temperature typical of a filament.

As mentioned in Sect.~3.2, we analyzed the kinematics of four blobs located at various altitudes in the EP. We found interesting to connect the prominence kinematics with the physical conditions in this falling material which contributes to the reformation of the prominence. We computed the variation (with altitude or time) of the blob's He~{\sc ii}~304~\AA\ emission with a careful subtraction of the (varying) background and a normalisation by the quiet Sun He~{\sc ii} intensity, known to be of the order of 8\,400~ergs~s$^{-1}$~cm$^{-2}$~sr$^{-1}$
\citep{1978ApJS...37..485V}. The ratio I(blob)/I(QS) was found to vary between 0.1 (mostly at high altitudes ~100\,000~km) to more than 1 (at low altitudes ~20\,000~km ), with no evident trend with time. Although we do not have spectral information, one can derive some information from the non-LTE modellisation, such as performed by \citet{2001A&A...380..323L} for prominences. Their range of temperatures is limited to $\sim$20 000 K but their range of widths (up to $\sim$10\,000~km) is well comparable to the size of our blobs (between 4\,000 and 8\,400~km, assuming a spherical shape). From their figures 15 and 19, one can see that the He~{\sc ii}~304~\AA\ intensity is of the order of 1\,000 ($\pm$ 200)~ergs~s$^{-1}$~cm$^{-2}$~sr$^{-1}$ which implies a mean I(prominence)/I(QS) ratio of 0.12. From this result, one can derive that the I(blob)/I(QS) values found at high altitudes correspond to a low temperature, at least in the range of values (6\,000--20\,000 K) of \citet{2001A&A...380..323L}. The high values found at lower altitudes can be tentatively interpreted as the signature of much higher temperatures (corresponding to the usual 70 000 K value).

\section{Discussion}
\label{sect4}
Although the velocity evolution profile of the prominence/filament analysed here is similar to torus-ring eruptions, a very clear twist of the prominence body  above a critical value of 2${\pi}$ which evolves into a writhe does favour the kink instability scenario. However, a kink instability may not have been the only trigger for this eruption but, for instance,  the data do not reveal any indication of magnetic reconnection.  A clear signature of the VERY triggering
of destabilization based on the present observations
is, therefore, difficult to establish. Conclusive observations
on the onset of prominence eruptions would require magnetic field measurements in the
footpoints of eruptive prominences, combined with
imaging and spectroscopic observations.

We examined  the possible  role of  kink instability in the eruption of the prominence discussed here. A magnetic flux rope  becomes kink-unstable if the twist of the magnetic  field  $T_{w}$, a measure of the number of  turns of the magnetic field lines about the prominence main axis, exceeds a critical value of $2\pi$ \citep{1981GApFD..17..297H}.
During the instability  a flux tube would  helically untwist reducing the magnetic stress due to  this  instability.  The magnetic helicity, $H_m$, can be expressed as the combination of twist and writhe, i.e. $H_m = \Phi^{2}(T_{w} + W_{r})$, where $\Phi$ is the axial magnetic flux in the flux rope  and $W_r$ is the  measure of the twist of the rope around itself. Because of the conservation of magnetic helicity  \citep{1984GApFD..30...79B}, the twist will transfer  into a writhe, $W_r$,  i.e. forming a ribbon-like structure.  \citet{1991SoPh..136..151V} found that non-eruptive prominences have a total twist less than $2\pi$ (less than one complete turn) and, on the contrary, all eruptive prominences that had a total twist $\geq 2.5\pi$ (one and a quarter turns) erupted. On theoretical grounds \citet{1990ApJ...361..690M} found that a purely axial loop which is  twisted by its footpoints  becomes kink unstable  when the twist exceeds 4.8$\pi$.

In the present case the prominence shows a left-hand twist of 6$\pi$ which during the eruption transfers from the prominence legs to its upper body forming  a counter-clockwise writhe. The left-hand twist of the prominence body can be clearly seen on the AIA He~{\sc ii} 304~\AA\ images while the writhe of the eruptive  filament  is evident from  the images of EUVI/STEREO B (Fig.~\ref{fig3}). A flux rope helical  twisting and  writhing of the same sign is a basic condition for  kink instability to work (Hood and Priest 1979). That makes us to suggests  that the eruptive prominence analysed here was possibly submitted to a kink instability, although other instabilities may also have played some
role.

The eruptive prominence was located in the northern periphery of a large-scale CME magnetic configuration. Therefore, the background magnetic field is asymmetric with respect to  the filament position. There are several observational \citep[e.g.][]{2003ApJ...595L.135J, 2006ApJ...653..719A, 2007SoPh..246..365G, 2009ApJ...696L..70L, 2011RAA....11..594S} and theoretical \citep{2005ApJ...630L..97T, 2006PhRvL..96y5002K, 2007ApJ...668.1232F, 2008ApJ...679L.151L} studies suggesting a confining role of the overlying magnetic field during failed filament eruptions due to a slowly decreasing gradient of the magnetic field and a strong field intensity at low altitude. Recently, \citet{2009ApJ...696L..70L} investigated a failed filament eruption in which two filaments are asymmetrically located in coronal loops. The authors   compared the confinement ability of symmetric and asymmetric fields and found that the magnetic confinement of an asymmetric field is stronger than that of a symmetric one. \citet{2011RAA....11..594S} made such calculations for six EPs, from which five  represented failed eruptions and one a successful eruption. They suggested that an asymmetric background field is an important factor leading to failed filament eruptions.

\section{Conclusions}

We examined  the kinematic and  helicity pattern together with the morphological and geometrical evolution of an EP, based on  He~{\sc ii}~304~\AA\ AIA/SDO and EUVI/STEREO B observations. The unique combination of high-resolution limb observations of the EP in  AIA and  a central meridian position in  EUVI B permitted a detailed analysis of the prominence eruption on 2010 March 30. The obtained results can be summarised as follow:

-- The EP appeared as a helically twisted MFR with fixed footpoints. The prominence body was composed of  left-hand twisted threads around the main prominence axis.
The twist  during the eruption was estimated at $6\pi$ (3 turns).

-- The EP  twist progressively converted into a  left-hand writhe that is well traced in the EUVI/STEREO B images. The position of the crossing-point of the writhed prominence body descended while the eruption was progressing. Co-temporally the drain of prominence plasma towards the chromosphere can be followed in the AIA/SDO images.

--  The height-time profile of the EP revealed  three distinctive phases of prominence eruption: a prominence activation phase, accelerating phase, and eruptive phase, the last one with constant velocity.  Consequently the prominence
contracted to its primary  location after reaching a maximum height of 526 Mm. The prominence/filament partially reformed suggesting that either part of the material has drained into the photosphere or not all of the prominence material
returned to its primary temperature.

-- The draining plasm a blobs at low altitudes were found to have high temperatures which could result from the encounter of the downflowing plasma with the static plasma in the legs and the resulting heating compression.

-- The EP was associated with a narrow expanding loop and an empty cavity underneath located in the northern periphery of a CME which appeared at the eastern limb and was triggered by the prominence eruption.

\begin{acknowledgements}
The authors thank the anonymous referee for the  helpful suggestions and comments.  The AIA data are courtesy of SDO (NASA) and the AIA consortium. Research at Armagh Observatory is grant-aided by the
N. Ireland Department of Culture, Arts and Leisure. This work is grant-aided by the Bulgarian Academy of Sciences via the Institute of Astronomy.
The authors thank the STEREO/SECCHI consortium for providing the data. The
SECCHI data used here were produced by an international consortium of the Naval
Research Laboratory (USA), Lockheed Martin Solar and Astrophysics Lab (USA), NASA
Goddard Space Flight Center (USA), Rutherford Appleton Laboratory (UK), University
of Birmingham (UK), Max-Planck-Institut for Solar System Research (Germany), Centre
Spatiale de Li\`ege (Belgium), Institut d'Optique Theorique et Appliqu\'ee (France), Institut
d`Astrophysique Spatiale (France). JCV and EB acknowledge the support of C.N.R.S. in the frame of the CNRS/BAS
agreement ( 28 April 2010, \# 66459). They also acknowledge the support of
C.N.E.S. (Space French Agency) for easing the access to AIA/SDO data. They thank Claude
Mercier for her continuous support.
\end{acknowledgements}

\bibliographystyle{aa}







\end{document}